\documentclass[sigconf]{acmart}

\AtBeginDocument{%
  }

\newcommand{\up}[1]{\small ($\textcolor{green}{\blacktriangle}#1\%)$}

\newcommand{\name}{\textsc{FactIR}}
\setcopyright{acmlicensed}
\copyrightyear{2018}
\acmYear{2018}
\acmDOI{XXXXXXX.XXXXXXX}

\acmConference[Conference acronym 'XX]{Make sure to enter the correct
  conference title from your rights confirmation emai}{June 03--05,
  2018}{Woodstock, NY}
\acmISBN{978-1-4503-XXXX-X/18/06}

\copyrightyear{2025}
\acmYear{2025}
\setcopyright{cc}
\setcctype{by}
\acmConference[WWW Companion '25]{Companion Proceedings of the ACM Web
Conference 2025}{April 28-May 2, 2025}{Sydney, NSW, Australia}
\acmBooktitle{Companion Proceedings of the ACM Web Conference 2025 (WWW
Companion '25), April 28-May 2, 2025, Sydney, NSW, Australia}
\acmDOI{10.1145/3701716.3715300}

\begin{document}

\title{FactIR: A Real-World Zero-shot Open-Domain Retrieval Benchmark for Fact-Checking}


\author{Venktesh V}
\affiliation{%
 \institution{Factiverse AI and Delft University of Technology}
 \city{Delft}
 \country{Netherlands}}
 \email{V.Viswanathan-1@tudelft.nl}
 
\author{Vinay Setty}
\affiliation{%
  \institution{Factiverse AI and University of Stavanger}
  \city{Stavanger}
  \country{Norway}
  }
\email{vsetty@acm.org}

\renewcommand{\shortauthors}{Trovato et al.}

\begin{abstract}
The field of automated fact-checking increasingly depends on retrieving web-based evidence to determine the veracity of claims in real-world scenarios. A significant challenge in this process is not only retrieving relevant information, but also identifying evidence that can both support and refute complex claims. Traditional retrieval methods may return documents that directly address claims or lean toward supporting them, but often struggle with more complex claims requiring indirect reasoning. While some existing benchmarks and methods target retrieval for fact-checking, a comprehensive real-world open-domain benchmark has been lacking. In this paper, we present a real-world retrieval benchmark \name{}, derived from Factiverse production logs, enhanced with human annotations. We rigorously evaluate state-of-the-art retrieval models in a zero-shot setup on \name{} and offer insights for developing practical retrieval systems for fact-checking. Code and data are available at \url{https://github.com/factiverse/factIR}.
\end{abstract}

\ccsdesc[500]{Information systems~Information retrieval}

\keywords{Fact-checking, Retrieval benchmark}

\maketitle

\section{Introduction}
The rapid spread of information through digital media has increased the need for accurate, automated fact-checking to combat misinformation. Automated fact-checking systems verify claims by retrieving relevant evidence from vast, often unstructured web data~\cite{guo2022survey,schlichtkrull2023averitec}. A key challenge lies in the retrieval component, which traditionally prioritizes similarity. However, as misinformation becomes more complex, verifying multifaceted claims often requires reasoning from indirectly related evidence. For instance, validating a claim about vaccine safety may involve documents on production processes, clinical trials, or regulatory approvals. This complexity highlights the need for retrieval methods and benchmarks that go beyond traditional approaches.

Existing fact-checking benchmarks fall short of addressing real-world retrieval challenges. Popular datasets like FEVER~\cite{thorne-etal-2018-fever} and FEVEROUS~\cite{aly2021feverous} rely on synthetic claims, with retrieval limited to Wikipedia. While recent works such as ClaimDecomp~\cite{claimdecomp}, QABriefs~\cite{QABriefs}, and AVeriTeC~\cite{schlichtkrull2023averitec} incorporate fact-checks from professional fact-checkers, they constrain retrieval to pre-verified justification documents or lack dedicated retrieval corpora. This highlights the need for benchmarks that better simulate open-domain, real-world retrieval scenarios critical for advancing fact-checking systems.

In this paper, we present a novel retrieval benchmark rooted in real-world fact-checking data to address existing gaps. Our benchmark is derived from Factiverse production logs, incorporating evidence retrieval data and human annotations from an operational fact-checking system~\cite{10.1145/3626772.3657663,10.1145/3626772.3661361}. By leveraging data collected under genuine production conditions, we aim to provide a benchmark that accurately represents the unpredictable and multifaceted nature of real-world fact-checking. This dataset allows us to evaluate how retrieval models handle diverse information sources, incomplete knowledge contexts, partially relevant evidence, and claims that require subtle reasoning beyond simple factoid matching. Moreover, as most existing fact-checking systems rely on off-the-shelf retrievers pre-trained on datasets from other domains and used in a zero-shot setting~\cite{sriram-etal-2024-contrastive,gupta-etal-2022-dialfact}, it becomes essential to assess their generalization performance under these realistic constraints, where fine-tuning is often infeasible. In this paper, we adopt the same zero-shot setting to ensure our evaluation aligns with real-world usage scenarios.

Our contributions are threefold. First, we provide a real-world retrieval benchmark based on actual usage logs and human annotations, filling a critical gap in fact-checking research. Second, we conduct an extensive evaluation of state-of-the-art retrieval models (zero-shot), examining their ability to handle complex claims in authentic conditions. Finally, we offer insights and recommendations for designing retrieval systems that are not only effective in controlled settings but robust enough for real-world fact-checking, where the retrieval process must often synthesize indirect and inferential evidence. Through our work, we aim to foster the development of fact-checking systems that can better meet the challenges of today’s information ecosystem, supporting fact-checkers and automated systems alike in their mission to mitigate misinformation.
\section{Related Work}

Over the years, a range of benchmarks have been introduced to advance research in automated fact-checking. Early benchmarks such as FEVER~\cite{thorne-etal-2018-fever} and FEVEROUS~\cite{aly2021feverous} have been widely adopted, providing large-scale datasets for claim verification. These benchmarks include retrieval corpora but are based on synthetic claims generated from Wikipedia content, limiting retrieval to a constrained domain of Wikipedia articles. While these datasets have driven significant progress, they fail to reflect the complexity and diversity of real-world claims and evidence.

More recently, datasets like ClaimDecomp~\cite{claimdecomp} and QABriefs~\cite{QABriefs} have emerged, leveraging real-world fact-checks conducted by professional fact-checkers. These datasets introduce claim decomposition, where complex claims are broken down into sub-questions to facilitate evidence retrieval and verification. However, their retrieval process remains limited to pre-verified justification documents, which simplifies the task and does not simulate realistic retrieval challenges. Similarly, AVeriTeC~\cite{schlichtkrull2023averitec} incorporates human-authored questions paired with search-engine-retrieved answers, offering a step toward open-domain retrieval. Nevertheless, it lacks a dedicated retrieval corpus, restricting its ability to evaluate retrieval performance comprehensively.

Other benchmarks, such as QuanTemp~\cite{10.1145/3626772.3657874}, focus on numerical fact-checking and their primary focus is on numerical reasoning based verification, and retrieval is not directly evaluated. 

Despite these advancements, current benchmarks either rely on constrained retrieval settings or pre-verified evidence, failing to fully capture the challenges of open-domain, real-world retrieval where evidence must be identified from diverse, unstructured sources. Addressing these gaps is the primary goal of this paper.
\begin{figure}
    \centering
    \includegraphics[width=0.9\linewidth]{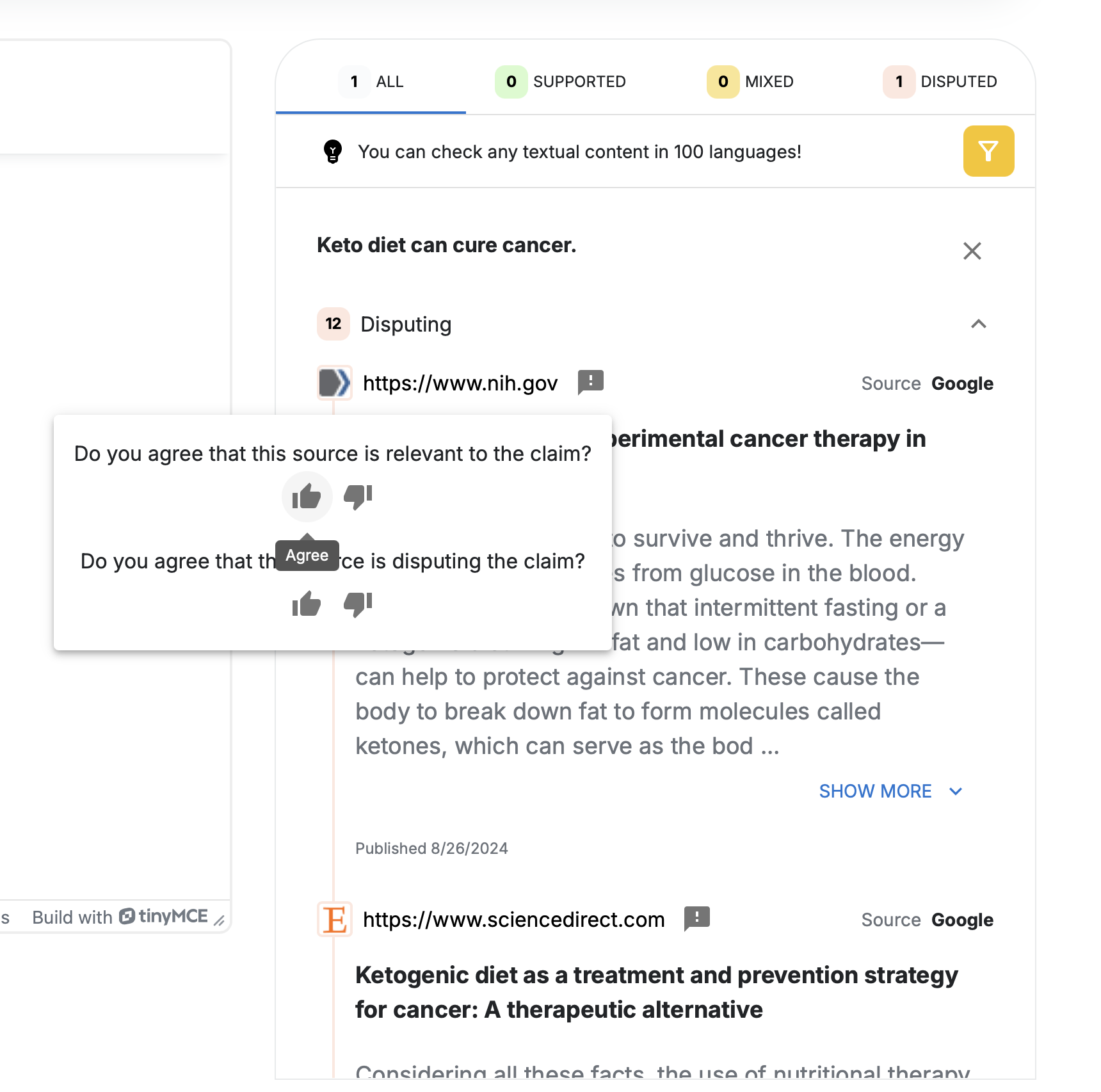}
    \caption{Factiverse fact-check editor with feedback mechanism. Example for the claim ``Keto diet can cure cancer.''}
    \label{fig:fceditor}
    \vspace{-10pt}
\end{figure}

\section{Data Curation}
We employ the Factiverse Fact-Check Editor~\cite{10.1145/3626772.3657663}, a web-based tool designed for automated fact-checking with a built-in feedback mechanism. As shown in Figure \ref{fig:fceditor}, users input claims they wish to fact-check, and the system aggregates evidence retrieved from multiple search engines, such as Google and Bing, ensuring comprehensive coverage of web content. The editor utilizes query generation and evidence filtering to retrieve the most relevant documents~\cite{10.1145/3626772.3661361,10.1145/3627673.3679985}. Users can then provide detailed feedback on two aspects: the relevance of the evidence to the claim and the correctness of the stance, whether the claim is supported or refuted.

To ensure high-quality labels for this benchmark, we focus on feedback collected from domain experts, specifically professional fact-checkers and individuals with journalism backgrounds, who verify claims using Factiverse’s production deployment. The claims are organically generated by users as they engage in tasks covering a variety of topics, including politics, healthcare, and the economy. The feedback was collected over the period spanning 2021 to 2023.

The annotators were provided with the following instructions and examples: \textbf{Evidence is deemed relevant to a claim if the extracted snippet can help verify the claim’s veracity.} For example, for the claim ``Keto diet can cure cancer,'' a document discussing the general benefits of the ``keto diet'' without addressing its effect on cancer would be considered irrelevant. On the other hand, a scientific study examining the ``ketogenic diet as a treatment and prevention strategy for cancer'' (as illustrated in Figure~\ref{fig:fceditor}), even if it disproves the claim, would be considered relevant.

\begin{figure}
\begin{minipage}{0.22\textwidth}
\captionof{table}{Topical distribution}
\label{tab:topical}
\begin{tabular}{lr}
\hline
\textbf{Topic} & \textbf{Count}  \\

\midrule
Politics & 26 \\
Economy & 24 \\
Health & 21 \\
Law & 6 \\
Climate & 8 \\
Education & 3 \\
Other &  12\\

\bottomrule
\end{tabular}\end{minipage}
\begin{minipage}{0.22\textwidth}
\captionof{table}{Numerical  taxonomy}
\label{tab:taxonomy}
\begin{tabular}{lr}
\hline
\textbf{Topic} & \textbf{Count}  \\

\midrule
Statistical & 40 \\
Temporal & 27 \\
Comparative & 8 \\

\bottomrule
\end{tabular}
\end{minipage}

\end{figure}

\begin{table*}[!ht]
    \centering
    \small
    \begin{tabular}{lccccccccc}
    \toprule
     \textbf{Method}& \multicolumn{1}{c}{nDcG@5} &\multicolumn{1}{c}{Recall@5} &\multicolumn{1}{c}{nDcG@10} & \multicolumn{1}{c}{Recall@10} & \multicolumn{1}{c}{nDcG@100} & \multicolumn{1}{c}{Recal@100} \\
     \midrule

    \midrule
    
    \textbf{Lexical} & & \\
            BM25 \cite{bm25} & 0.288 & 0.253 &0.345 & 0.421& 0.475&0.779 \\

     \midrule
     
      \textbf{Sparse} & & \\
            SPLADEV2 \cite{SPLADEv2} & 0.287 & 0.231& 0.336 & 0.384 & 0.473 & 0.783 \\
     \midrule
      \textbf{Dense} & & \\
            Stella-en-v5 & 0.090 & 0.065 & 0.098 & 0.110 & 0.158 & 0.314 \\
            DPR \cite{karpukhin-etal-2020-dense} & 0.247 &  0.219 & 0.292 & 0.356 & 0.404 & 0.670    \\
            ANCE \cite{ance} & 0.246 & 0.184  & 0.289 & 0.327 & 0.419 & 0.691\\

                        tas-b \cite{tas-b} & 0.289 & 0.232 & 0.336 & 0.399 & 0.468 & 0.771 \\
                        MPNet \cite{mpnet} & 0.290 & 0.250 & 0.327& 0.388 & 0.464& 0.767 \\
       Contriever \cite{contriever} & 0.299& 0.249 & 0.346 & 0.406 & 0.471 & 0.760\\
    COlBERTV2 \cite{santhanam-etal-2022-colbertv2} & 0.252 &  0.230 & 0.325 & 0.419 & 0.465 & 0.789  \\
    Snowflake-arctic-embed-s \cite{merrick2024embeddingclusteringdataimprove} & \textbf{0.367} \up{27.43} & \textbf{0.302} \up{19.37} & \textbf{0.420} \up{21.74} & \textbf{0.480} \up{14.01} & \textbf{0.529} \up{11.37} & \textbf{0.795} \up{2.05} \\
\midrule
\textbf{Re-Ranker} \\
BM25 + & \\
 - ColBERTV2 &0.265 & 0.253 &  0.333 & 0.424 & 0.464 & 0.759 \\
- MARCO-MiniLM-H384 & 0.293 & 0.247     & 0.349 & 0.408 & 0.485 & 0.779\\

 - MARCO-MiniLM-en-de & 0.278 & 0.264 & 0.342 & 0.426 & 0.479  & 0.779 \\
- bge-reranker-base & 0.222 & 0.205 & 0.301 & 0.398 & 0.452 & 0.779\\
- bge-ranker-v2 & 0.252 & 0.235 & 0.312 & 0.399 & 0.460 & 0.779 \\
- Jina-reranker-v2 & 0.270 & 0.245 & 0.336 & 0.421 & 0.474 & 0.779\\
- gte-multilingual & \underline{0.308} \up{6.04}& \underline{0.277} \up{9.49} & \underline{0.368} \up{6.67} & \underline{0.437} \up{3.80}& \underline{0.496} \up{4.42} & \underline{0.779} \\


\midrule


    \end{tabular}
    \caption{Retrieval results on \name{}, nDCG@10 across datasets. The best results are in bold and the second best results are underlined with \% improvements indicated by \up{} over the baseline BM25 being specified in brackets. }
     \vspace{-1em}
    \label{tab:main_result}
\end{table*}
\vspace{-0.6em}

\subsection{Benchmark Statistics}
 The benchmark comprises, \textbf{1413} claim-evidence  pair relevance annotations with a total of \textbf{90047} documents in the corpus collection and 100 claims with an average of \textbf{13.89} documents / relevance assesments per query. We provide fine-grained topical analysis in Table \ref{tab:topical}. We observe that a number of claims in the benchmark are \textit{quantitative or temporal} in nature, as shown in Table \ref{tab:taxonomy}. We also observe that many claims are \textit{compositional} comprising multiple aspects, making \name{} a challenging retrieval benchmark. We performed a qualitative meta-analysis of relevance assignments with the help of two researchers who were asked to annotate as ``1" when they deem the relevance assessment to be correct else ``0". The annotators found the relevance assessments to be of high quality (\textbf{88.03\%} was deemed to be correct) with a high agreement of \textbf{0.946} as indicated by Cohen's kappa.

\vspace{-0.7em}
\section{Experimental Setup}
 We evaluate a range of state-of-the-art retrieval and re-ranking approaches on \name{},  with two V100S-PCIE-32GB GPUs. 

\textbf{Lexical and Sparse retrieval}: We employ ElasticSearch's BM25 implementation due to low latency and ease of use~\cite{fogelberg2023search}. We also employ SPLADEV2 \cite{SPLADEv2} which is a neural model employing query and document sparse expansions.

\textbf{Dense retrieval}: DPR ~\cite{karpukhin-etal-2020-dense} comprises a query, document bi-encoder model, and we employ the open source model trained on multiple Question Answering datasets \textit{facebook-dpr-question-encoder-multiset-base}. We also include ANCE~\cite{ance} in our evaluation, which is a bi-encoder model that samples hard negatives through Nearest Neighbour search over the corpus index, yielding better negatives. We employ \textit{msmarco-roberta-base-ance-first} trained on MS-MARCO \cite{bajaj2018ms}. Tas-b \cite{tas-b} is a bi-encoder trained using supervision from a cross-encoder. We benchmark the recent \textit{snowflake-arctic-embed-s} \cite{merrick2024embeddingclusteringdataimprove} which improves contrastive pre-training through semantic clustering-based source stratification. 

\textbf{Late-Interaction Models}: We implement the late-interaction model ColBERTv2 \cite{colbert,santhanam-etal-2022-colbertv2} in \name{}. It is a late-interaction model that employs a cross-attention-based MaxSim operation betweeen query and document token representations.

\textbf{Re-rankers:} We employ several state-of-the-art cross-encoder models including Large Language Models (LLMs) for re-ranking. We employ ColBERTV2 in re-ranker mode due to better relevance estimate from late-interaction. We evaluate BERT based cross-encoders like \textit{cross-encoder/mmarco-mMiniLMv2-L12-H384-v1} and \textit{cross-encoder/msmarco-MiniLM-L12-en-de-v1} that have shown to achieve impressive performance on BEIR \cite{beir} benchmark. We also benchmark more recent re-rankers like \textit{jinaai/jina-reranker-v2-base-multilingual} which employs flash-attention. We also evaluate LLM based re-rankers like \textit{Alibaba-NLP/gte-multilingual-reranker-base}

\textbf{Metrics:}  We choose Normalized Discounted Cumulative Gain (nDCG@k) and Recall@k as the primary metrics for our results. 
\vspace{-1.2em}

\section{Results and  Analysis}
In this section, we analyze how retrieval and re-ranking approaches perform on the \name{} benchmark (Table \ref{tab:main_result}). 

\textbf{Lexical and Sparse retrievers}: We observe that lexical models like BM25 and learned sparse retrievers like SPLADEV2 perform surprisingly well on \name{} benchmark, as shown in Table \ref{tab:main_result}. They outperform several dense retrievers, proving to be strong baselines.

\textbf{Classical dense retrieval models fail on OOD data}: Since models like DPR, Contriever are frequently used in a zero-shot fashion for fact-checking tasks \cite{gupta-etal-2022-dialfact,sriram-etal-2024-contrastive}, we evaluate their performance on \name{}. We observe that all dense retrievers have sub-par performance when compared to BM25. This could primarily be caused by domain shift from pre-training data to claims in \name{}. Additionally, these models might also be incapable of handling such complex real-world claims as models like DPR are trained on datasets mostly containing factoid queries \cite{sriram-etal-2024-contrastive}.  We observe that DPR which has been trained on QA datasets performs worse than other dense retrievers like Contriever or tas-b which are pre-trained on MS-MARCO. We observe that Contriever is closer in performance compared to BM25 compared to other dense retrievers.

\textbf{Retrieval models with strong stratification based training objectives generalize better}:
From Table \ref{tab:main_result}, we observe that snowflake-arctic-embed-s outperforms lexical sparse and other state-of-the-art dense retrieval models  as measured by nDCG@k and Recall@k demonstrating better generalization performance. We hypothesize that this is primarily due to the semantic clustering based source stratification \cite{merrick2024embeddingclusteringdataimprove} based training objective. This objective is inspired by the cluster hypothesis \cite{clustering_hypothesis} and further builds upon the topic aware sampling philosophy of tas-b \cite{tas-b} and the hard negative mining aspect of ANCE \cite{ance} leading to superior performance. We posit that this training regime causes \textit{snowflake-arctic-embed-s} to retrieve topically and contextually relevant evidence for the claim while ignoring topically similar hard negatives.

\textbf{LLM-Based Re-Rankers Show Superior Generalization}: After retrieving the top-100 documents using BM25, we re-rank them with a variety of models, as shown in Table \ref{tab:main_result}. Among these, the LLM-based re-ranker \textit{gte-multilingual-reranker-base} outperforms other cross-encoder models pre-trained on MSMARCO. We attribute this performance to the architecture of mGTE and the extensive pre-training of \textit{gte-multilingual} \cite{zhang-etal-2024-mgte} on data sourced from a wide range of tasks and domains, reducing source bias. This multi-domain pre-training enhances its generalization capabilities, enabling it to perform effectively across diverse retrieval scenarios.

\vspace{-1em}
\section{Conclusion}
In this work, we introduce \name{} for benchmarking of retrieval for fact-checking in an open-domain setup. \name{} is a collection of high quality real-world claims along with high quality relevance assessments from users. We evaluate a wide range of retrieval and re-ranking models and observe lexical and sparse retrievers to be strong baselines. We also observe that strong training objectives lead to impressive generalization performance. We also release our library which is easily extendable to new retrievers and re-rankers.
\section{Acknowledgements}
This work is partly funded by the Research Council of Norway project EXPLAIN (grant no:
337133).

\bibliographystyle{ACM-Reference-Format}
\bibliography{references}


\end{document}